\documentclass[a4paper,11pt]{article}
\usepackage{pos}

\title{The Role of the Heliosphere in Shaping the Observed Cosmic Ray Spectral Anisotropy}
\ShortTitle{The Role of the Heliosphere in the Observed Cosmic Ray Spectral Anisotropy}

\author*[a]{Vanessa L\'opez-Barquero}
\author[b]{Andr\'es Mar\'in Portuguez}
\author[c]{Paolo Desiati}
\author[c]{Juan Carlos D\'iaz-V\'elez}

\affiliation[a]{Department of Astronomy, University of Maryland,\\
  College Park, MD 20742-2421, USA}

\affiliation[b]{Escuela de F\'isica, Universidad de Costa Rica, \\
San Jos\'e 11501-2060, Costa Rica.}

\affiliation[c]{Wisconsin IceCube Particle Astrophysics Center (WIPAC), University of Wisconsin,\\
Madison, WI 53703, USA}

\emailAdd{vlopezb@umd.edu, vanelopezb@gmail.com}

\abstract{Experimental results by Milagro, HAWC, and ARGO-YBJ have observed variations in the energy spectrum of cosmic rays at TeV scales in different regions of the sky. These findings on the spectral anisotropy provide insights into cosmic ray behavior. This work explores the impact of galactic cosmic ray interactions with the heliosphere in creating the observed spectral anisotropy features. Specifically, the features around 1-10 TeV, where our previous studies on the heliosphere have shown the greatest effects. In this project, we integrate particle trajectories in a state-of-the-art MHD-kinetic heliosphere model that includes the effects of the solar cycle and interaction with the interstellar medium’s magnetic field. With these elements, this is the first time the exact effects of the heliosphere’s magnetic field are tested to determine their influence on galactic cosmic rays and their spectral anisotropy. In our results, we identified an area on the map that exhibits a distinct cosmic ray energy spectrum compared to the all-sky distribution. This area approximately coincides with Region A, where observations have found a harder energy spectrum than the isotropic spectrum.
}

\ConferenceLogo{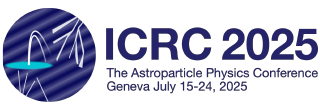}

\FullConference{39th International Cosmic Ray Conference (ICRC2025)\\
 15–24 July 2025\\
Geneva, Switzerland\\}

\begin{document}
\maketitle

\section{Introduction} \label{sec:intro}

Observations by the Milagro, ARGO-YBJ, and HAWC  experiments \cite{Milagro2008, Argo2013, HAWC2014} have revealed deviations from spatial isotropy in the TeV cosmic-ray energy spectrum, offering new insights into particle propagation in our Galaxy. The Milagro water-Cherenkov telescope first reported "hot spots" in the northern celestial hemisphere, exhibiting relative intensity enhancements of order $10^{-4}$-$10^{-3}$ at multi‑TeV energies. The ARGO‑YBJ air-shower array corroborated these findings by mapping medium-scale excesses $\sim10^\circ$ with intensities up to $10^{-3}$ in the 10–20 TeV band. Most recently, the HAWC Observatory confirmed the presence of Milagro's two primary excess regions and identified a third enhancement consistent with ARGO‑YBJ's observations; these structures manifest at relative intensity levels as low as $10^{-4}$ on angular scales of order $10^\circ$. In terms of energy spectra, both Milagro regions' spectra are inconsistent with the isotropic cosmic-ray flux, with Region A modeled as hard-spectrum protons with a cutoff. HAWC further finds that the energy spectrum of Region A is distinct from the isotropic cosmic-ray spectrum, with Region A exhibiting a harder spectrum. These measurements collectively contribute to our understanding of spatial anisotropy in the TeV cosmic-ray sky and point towards the presence of energy-dependent spectral hardening in certain localized regions. 

As shown in our previous studies \cite{2025ApJ...983..106L,2017ApJ...842...54L}, we have seen that the heliosphere has a strong effect on galactic cosmic rays, especially the ones with rigidities on the order of 1-10 TV. Here, we explore the possibility that the heliosphere can impact the energy spectrum of particles and how such effects cause the observed spectral anisotropy features. 

\section{Heliospheric Magnetic Field Model} \label{sec:magfield}

\begin{figure}
    \centering
    \includegraphics[width=0.7\linewidth]{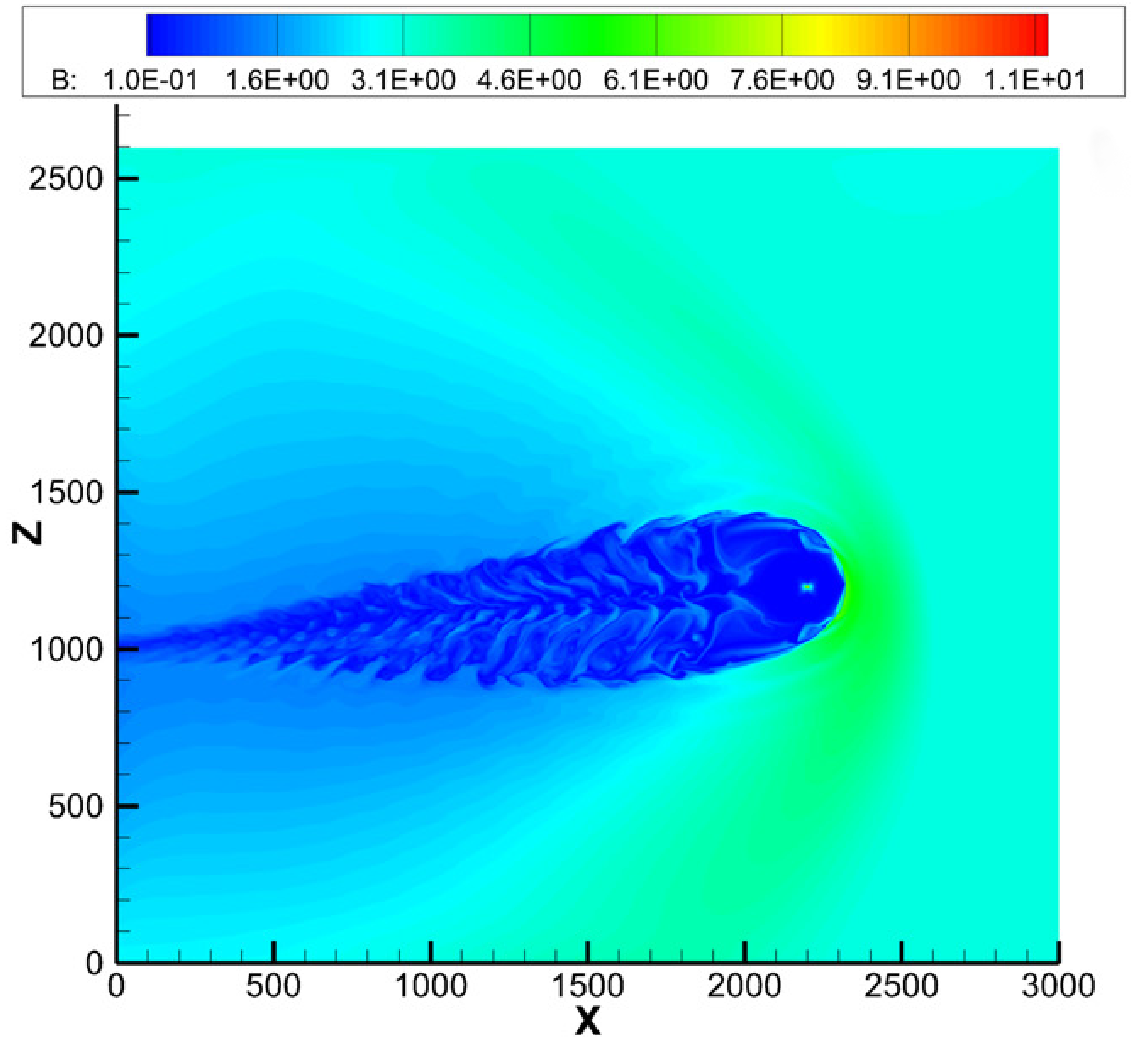}
    \caption{Model of the heliotail, as described by Pogorelov et al.(2015) \cite{Pogorelov_2015}, taking into account the effects of solar cycles. The units in the x and z directions are in astronomical units (AU), while the color indicates the strength of the magnetic field (\(|\mathbf{B}|\)) in microgauss ($\mu\text{G}$). The spatial extent of the tail along the x-axis is of a few thousand astronomical units (AU), while, in contrast, the extent in the z-direction is considerably more limited, spanning only a few hundred AU.}
    \label{F_heliotail_b_field}
\end{figure}

The numerical model used in this work to represent the heliosphere is described in \cite{Pogorelov_2015}. Figure~\ref{F_heliotail_b_field} shows the instantaneous $B$ field distribution in the heliotail simulation that takes into account solar-cycle effects. 
From the interaction of the solar wind with the local interstellar medium, the heliosphere is formed. Fundamental physical interactions, including charge exchange, photoionization, electron impact, recombination processes, and elastic collisions, establish crucial ties between ions and neutral atoms, and as a consequence, links between solar and interstellar material. These processes are responsible for the morphology of the heliosphere and the characteristics of the very local interstellar medium influenced by the heliosphere, which can extend over distances reaching thousands of astronomical units.
This approach employs an ideal Magneto-Hydrodynamic (MHD) treatment for ions and a kinetic multi-fluid model to analyze the behavior of neutral interstellar atoms as they enter the heliosphere.

\section{Particle Trajectory Integration} \label{sec:trajs}

As cosmic rays traverse a medium that possesses a magnetic field, they are subject to the Lorentz force:

\begin{equation}
    \mathbf{F} = q\left(\mathbf{E} + \frac{\mathbf{v}\times \mathbf{B}}{c}\right)
\end{equation}

In the case of the heliosphere and for the high-energy particles studied here, we can neglect the electric field E. We define particles using a 6-dimensional vector in phase space, $(\mathbf{r}, \mathbf{p})$, where $\mathbf{r}$ represents the particle's position and $\mathbf{p}$ denotes its relativistic momentum. For computational convenience, we set $c = 1$.

\begin{align}
    \frac{d\mathbf{p}}{dt} &= q\,\left(\mathbf{v}\times \mathbf{B}\right) \label{E_lorentz} \\ 
    \frac{d\mathbf{r}}{dt} &= \mathbf{v} \label{E_velocity_definition}
\end{align}

Propagation is achieved through the numerical integration of equations (\ref{E_lorentz}) and (\ref{E_velocity_definition}) in the context of a static magnetic field model of the heliosphere. At each time step, the Lorentz force acting on the particle is computed as a function of its instantaneous velocity and the magnetic field present at its specific position.

The algorithm implemented in the propagation code for this project is based on the Boris Push method \cite{boris1970relativistic}. This method offers significantly higher accuracy for a given time step compared to other integration algorithms \cite{Tretiak_2019}. The time steps for propagation are adjusted dynamically based on the strength of the magnetic field \cite{Desiati2014THETO}. This approach enhances the computational efficiency of the model.

Because of the extensive energy range of ground-based experiments, a trajectory dataset with a continuous energy distribution corresponding to rigidities from 30 GV to 300 TV was subjected to numerical integration. In total, $1.6\times10^{7}$ anti-proton trajectories were integrated backward in time, commencing from the Earth’s position, which is assumed to align with the Sun in the model's framework. The numerical integration process continues until the trajectories intersect with a sphere at a radius of 50,000 AU. The trajectories depict a mapping between a uniformly distributed particle ensemble on Earth and the corresponding distribution in the Local Interstellar Medium, as influenced by the local interstellar magnetic field and the heliospheric environment.

\section{Results} \label{sec:results}

\begin{figure}
    \centering
    \includegraphics[width=0.7\linewidth]{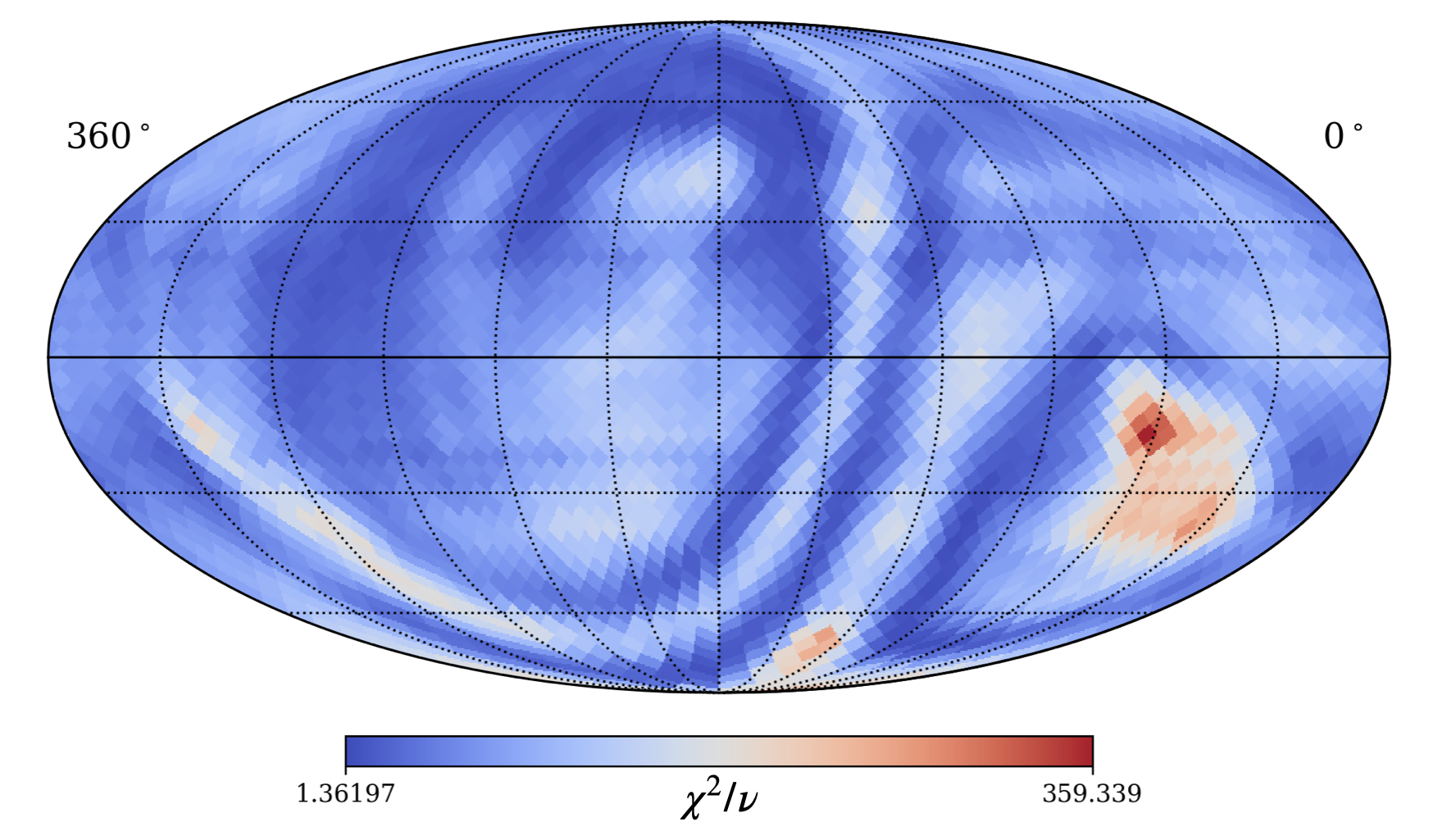}
    \caption{Cosmic Ray Spectral Anisotropy Map. This figure shows the reduced $\chi^2$ values for the spectral distribution of particles at a median energy of 7 TeV plotted in a map in equatorial coordinates. The red color in the map indicates areas where the spectral distribution differs from the overall distribution across the entire sky. This variation is attributed to interactions with the heliosphere. The red region in the Southern Hemisphere approximately corresponds to Region A in the HAWC data, where the energy spectrum is harder than the isotropic spectrum.}
    \label{fig:chi27tev}
\end{figure}

The different energy spectra were tested using the reduced $\chi^2$ test. This test will determine how likely the two spectra have the same distribution \cite{2010Gagunashvili}. This allows us to determine whether the heliosphere changes the energy spectrum. A specific region in the sky is compared with the distribution of the entire sky. This area is a 5-degree disc centered at each pixel in a HealPix \cite{2005Healpix} grid with Nside=16.

In Figure \ref{fig:chi27tev}, we present the results of our statistical test plotted on a map in celestial coordinates as detected on Earth. The areas denoted in red represent regions in the sky where the energy spectrum distribution differs from that of the entire sky. One feature to note is a red region on the Southern Hemisphere centered at $\sim$  ${45^\circ}$ longitude. This region is approximately aligned with Region A as identified in the HAWC data~\cite{HAWC2014}, where its energy spectrum is harder than that of the isotropic spectrum.

\section{Conclusions} 
\label{sec:conclusion}

Our results indicate that the heliosphere has a substantial impact on the energy distribution of particles. Different regions of the sky can be affected in distinct ways, with the spectral distribution varying based on the direction observed.

Thus, this study demonstrates how the interaction between galactic cosmic rays and the heliosphere contributes to the observed features of spectral anisotropy, particularly in the 1-10 TV range, where our prior research on the heliosphere has shown the most significant effects.

For the first time, we unveil the crucial role of the heliosphere's magnetic field in shaping the spectral anisotropy of galactic cosmic rays.


 



\bibliography{bibliography}{}
\bibliographystyle{aasjournal}


\end{document}